\documentclass[twocolumn,showpacs,preprintnumbers,amsmath,amssymb]{revtex4}
\input psfig.sty

\usepackage{color}
\usepackage{graphicx}
\usepackage{dcolumn}
\usepackage{bm}
\usepackage{latexsym}
\usepackage{epsfig}
\usepackage{multirow}
\usepackage{booktabs}
\usepackage{rotating}
\usepackage{bm}

\usepackage{color}
\usepackage{graphicx}
\usepackage{dcolumn}
\usepackage{bm}
\usepackage{epsfig}
\usepackage{graphicx}
\usepackage{dcolumn}
\usepackage{bm}
\usepackage[english]{babel}
\usepackage{booktabs}
\usepackage{multirow}

\newcommand{\be}{\begin{equation}}
\newcommand{\ee}{\end{equation}}
\newcommand{\beqq}{\setlength\arraycolsep{2pt}\begin{eqnarray}}
\newcommand{\eeqq}{\vspace{0cm} \end{eqnarray}}
\newcommand{\bea}{\begin{eqnarray}}
\newcommand{\eea}{\end{eqnarray}}

\newcommand{\mytilde}{\raise.17ex\hbox{$\scriptstyle\mathtt{\sim}$}}

\begin{document}

\title{Constraints on a possible evolution of mass density power-law index in strong gravitational lensing from cosmological data}

\author{R. F. L. Holanda$^{1,2,3}$} \email{holanda@uepb.edu.br}
\author{S. H. Pereira$^{4}$} \email{shpereira@feg.unesp.br}
\author{Deepak Jain$^{5}$} \email{djain@ddu.du.ac.in}

\affiliation{ $^1$Departamento de F\'{\i}sica, Universidade Federal de Sergipe, 49100-000, Aracaju - SE, Brazil,
\\ $^2$Departamento de F\'{\i}sica, Universidade Federal de Campina Grande, 58429-900, Campina Grande - PB, Brazil,\\$^3$Departamento de F\'{\i}sica Te\'{\o}rica e Experimental,\\ Universidade Federal do Rio Grande do Norte, 59300-000, Natal - RN, Brazil.\\$^4$Universidade Estadual Paulista (Unesp)\\Faculdade de Engenharia, Guaratinguet\'a \\ Departamento de F\'isica e Qu\'imica\\ Av. Dr. Ariberto Pereira da Cunha 333\\
12516-410 -- Guaratinguet\'a, SP, Brazil
\\$^5$Deen Dayal Upadhyaya College, University of Delhi, Sector 3, Dwarka, New Delhi 110078, India}



\begin{abstract}

In this work, by using strong gravitational lensing (SGL) observations along with Type Ia Supernovae  (Union2.1) and gamma ray burst data (GRBs), we propose a new  method to study  a
 possible  redshift evolution of $\gamma(z)$, the mass density power-law index of strong gravitational lensing systems. In this analysis, we assume the validity of cosmic distance duality
 relation and the flat universe. In order to explore the $\gamma(z)$ behavior, three  different parametrizations are considered, namely: (P1) $\gamma(z_l)=\gamma_0+\gamma_1 z_l$,
 (P2) $\gamma(z_l)=\gamma_0+\gamma_1 z_l/(1+z_l)$ and (P3) $\gamma(z_l)=\gamma_0+\gamma_1 \ln(1+z_l)$, where $z_l$ corresponds to lens redshift. If $\gamma_0=2$ and $\gamma_1=0$ the
 singular isothermal sphere model is recovered. Our method is 
performed on SGL sub-samples defined by different lens redshifts and velocity dispersions. For the
 former case, the results  are in full agreement with each other, while a 1$\sigma$ tension between the sub-samples with low ($\leq 250$ km/s) and high ($>250$ km/s)  velocity 
dispersions was obtained on the ($\gamma_0$-$\gamma_1$)  plane. By considering the complete SGL  sample, we obtain $\gamma_0 \approx 2$ and $ \gamma_1 \approx 0$ within 1$\sigma$ c.l. 
for all $\gamma(z)$ parametrizations.  However, we find the following best fit values of $\gamma_1$: $-0.085$, $-0.16$ and $-0.12$ for P1, P2 and P3 parametrizations, respectively, 
suggesting a mild evolution for $\gamma(z)$. By repeating the analysis with Type Ia Supernovae  from JLA compilation, GRBs and SGL systems this mild evolution is reinforced.

\end{abstract}

\maketitle

\section{Introduction}

Gravitational lensing phenomenon is one of the most successful predictions of the general relativity  theory  characterized by a bending of light when it passes close to a massive object. 
Particularly, two important quantities can be obtained from strong gravitational lensing (SGL) observations: the Einstein radius and time-delay distance. The former depends on the ratio of 
angular diameter distances (ADD) between lens/source and observer/source  while the second depends on three distances: the ADD between observer and lens, observer and source, and
 lens and source. Briefly, this effect is caused by the 
difference in length of the optical paths and the gravitational time dilation for the ray passing through the effective gravitational potential of the
 lens (Schneider, Ehlers \& Falco 1992; Kochanek, Schneider \& Wambsganss 2004). 

 Nowadays, SGL observations become a very important tool to measure cosmological parameters. For instance, each ADD in time-delay distance  is proportional to the inverse of
 Hubble constant, $H_0$. Actually, the possibility of independent determination of $H_0$ using time delay between images was suggested  in 1964 by Refsdal, however, only recently the
 technique has been competitive with other cosmological tests 
considering a flat $\Lambda$CDM scenario (Saha et al. 2006; Coe \& Moustakas 2009; Suyu et al. 2010, 2013). When combined with cosmic microwave background power spectrum, time-delay 
distance measurements are very effective at breaking degeneracies such as those between $H_0$ and $\omega$, the dark energy equation-of-state
 parameter  (see also  excellent reviews  in Kochanek, Schneider \& Wambsganss 2004 and Treu 2010). As a new approach, Paraficz \& Hjorth (2009)
 showed that the ADD to lens  can be obtained from a joint analysis between the gravitationally lensed quasar images and dispersion velocity of the lensing galaxy (see also Jee, Komatsu \& Suyu 2015 and Holanda 2016). 

The Einstein radius measurement is insensitive to Hubble constant since it is a ratio between two ADD. However, this quantity has been largely used to  constrain the cosmological parameters of several
 models {(Futamase \& Yoshida 2001, Biesiada 2006 and Grillo et al. 2008). An expressive work has been done recently by Cao et al. in which concerns applications of SGL 
data (Cao \& Liang 2011, Cao et al. 2015a, 2016a, 2016b) including 
statistical analyses of observed image separations (Cao \& Zhu 2012), lens redshifts (Cao et al. 2012) and more recently to test post-newtonian models of gravity at galaxy-scale
 (Cao et al. 2017a). (See also Mitchell et al. 2005 and Ofek et al. 2003 for additional applications).} SGL systems  were also used to constrain the cosmic equation of state 
parameter in XCDM cosmology and in the Chevalier - Polarski - Linder (CPL) 
parametrization, where $\omega$ is allowed
 to evolve with redshift as $\omega(z)=\omega_0 + \omega_1{z\over 1+z}$.  Particularly, Cao et al. (2015a) used 118 SGL systems from the Sloan Lens ACS Survey, 
BOSS emission-line lens survey, Lens Structure and Dynamics, and Strong Lensing Legacy Survey, improving the confidence regions on the parameter space. 
These authors also showed that the analyses with SGL may be complementary to type Ia Supernovae (SNe Ia) data. Very recently, SGL measurements have also been used 
jointly with SNe Ia observations to test the so-called cosmic distance duality relation (CDDR), $D_LD_A^{-1}(1+z)^{-2}=1$, where $D_A$ is the ADD and $D_L$ is the luminosity distance in a given redshift (Liao et a. 2016; Holanda, 
Busti \& Alcaniz 2016; Holanda, Busti, Lima \& Alcaniz 2016). No significant departure from the CDDR validity with this data set was verified.

However,  some  problems arise when one uses  SGL observations as cosmological tool, for instance, different values of $H_0$ are obtained
 from system to system. In this context,  Suyu et al. (2010, 2013) obtained a value of $H_0 = 70.6 \pm 3.1$ km/s/Mpc for B1608+656 system and $H_0 = 78.7^{+4.3}_{ -4.5}$ km/s/Mpc for RXJ1131–1231. 
Another important uncertainty source is the lens mass model, as different values of $H_0$ are obtained if one assumes either a singular isothermal spherical profile (SIS model, where $\rho \propto r^{-2}$) or  a
 spherically symmetric power-law mass distribution ($\rho \propto r^{-\gamma}$). The SIS profile has been widely used to describe lens galaxies, however, several studies have shown that
 the slopes of density profiles of individual galaxies show a non-negligible scatter from the SIS model (Koopmans 2005; Koopmans et al. 2009; Auger et al. 2010; Barnabe et al. 2010; Sonnenfeld et al. 2013). Moreover, by
 using 11 early-type galaxies, Ruff et al. (2011) found a mild evolution when  the $\gamma$ parameter was allowed to vary with redshift, which would indicate that dissipative processes  play some role 
in the growth of massive galaxies. In other words, a $\gamma$ evolution may play a crucial role on galaxy structures. This fact has been investigated  considering SGL observations and complementary 
probes in some cosmological scenarios, such as: $\Lambda$CDM, XCDM and $X(z)$CDM (Cao et al. 2015a; Li et al. 2016; Cui, Li \& Zhang 2017). By using
 a relation such as $\gamma(z)=\gamma_0 + \gamma_1 z$, no significant evidence for the evolution of $\gamma$ from SGL observation has been found. {Very recently, by taking the Planck's best-fitted 
cosmology, Cao et al. (2016a) considered SGL observations  and relaxed the assumption that stellar luminosity and total mass distribution follows the same power-law. Interestingly, they found that the presence of
 dark matter in the form of a mass component is distributed differently from the light (see also Schwab et al. 2010). Their results also suggested the need of treating low, intermediate 
and high-mass galaxies separately. At this point, it is very important to stress that the results of these previous studies were obtained by using some specific cosmological 
model in their analyses}\footnote{There are other important uncertainty sources in the SGL science, such as: velocity anisotropy, mass along the line of sight,  the mass-sheet 
degeneracy and  the environment of the lenses. However, in the present paper, we are considering only the mass profile shape.}. 

\begin{figure*}[ht]
\includegraphics[width=0.3\textwidth]{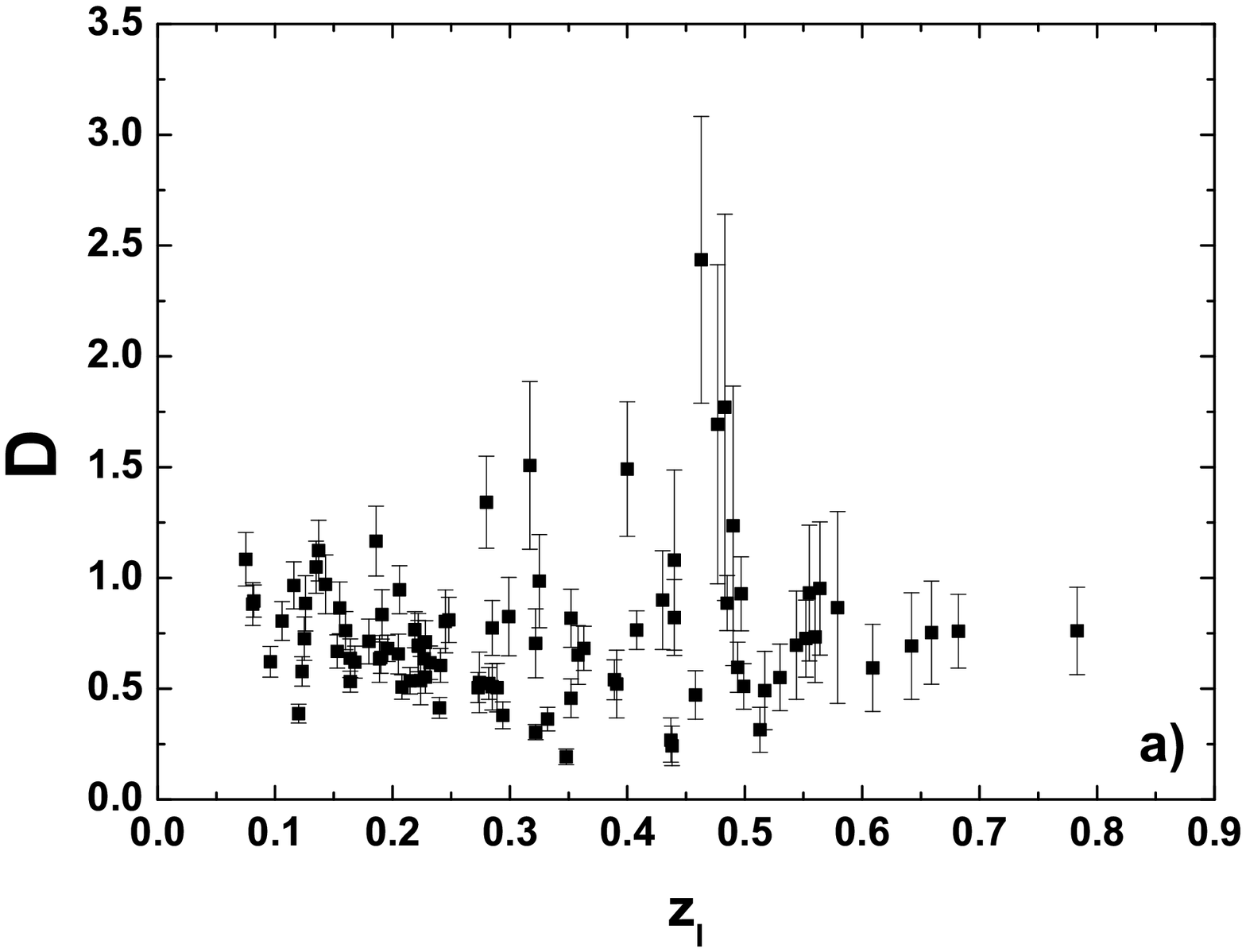}
\includegraphics[width=0.3\textwidth]{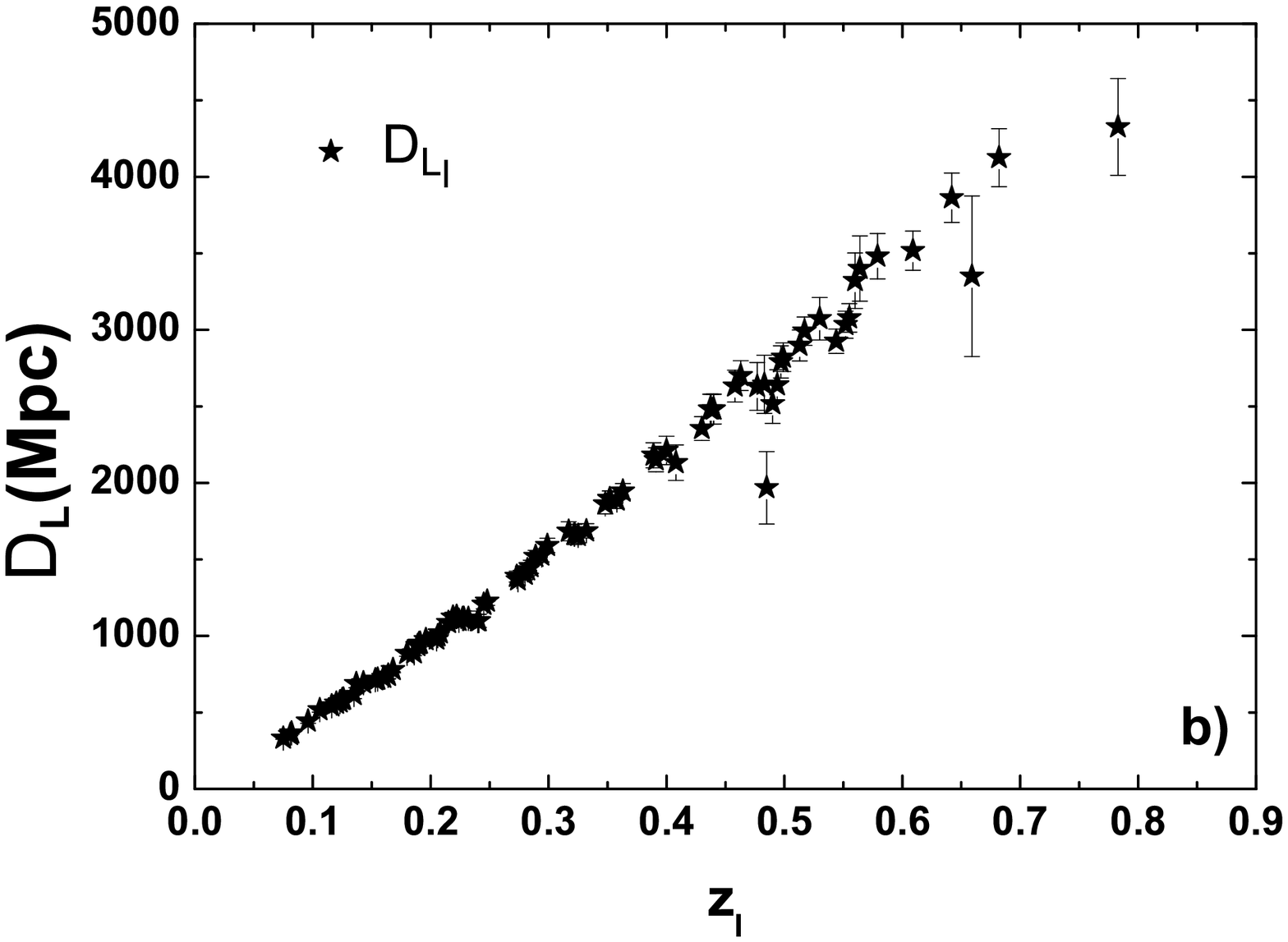}
\includegraphics[width=0.3\textwidth]{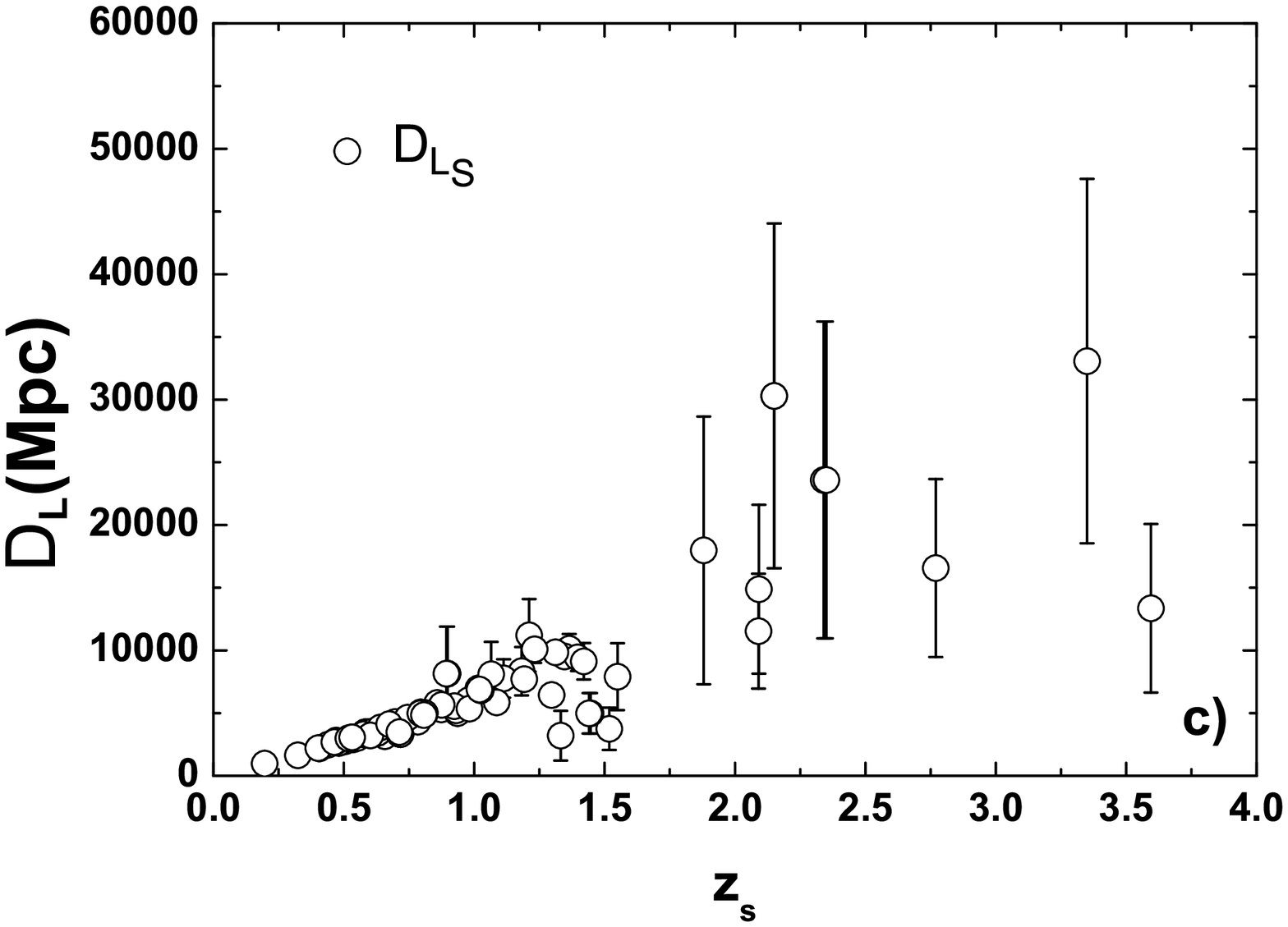}
\caption{ {Fig. (a) shows the complete SGL sample (92 points) used in our analyses (considering $\gamma_0=2$ and $\gamma_1=0$). The original SGL sample with 118 points can
 be found in Cao et al. (2015a). Figs. (b) and (c) show the luminosity distances to the lenses (filled star) and to the sources (open circles), respectively, of the 92 SGL 
systems in the Fig.(a). These luminosity distances were obtained from  original SNe Ia  (Suzuki et al. 2012) and gamma ray burst (Demianski et al. 2017) data. }}
\end{figure*}

{The main aim of this work is to perform  constraints on some $\gamma(z)$ parametrizations without explicitly using any cosmological model. From a theoretical point of view, only a flat universe  and  the 
validity of the CDDR relation are assumed. As data sets, we use SGL observations plus SNe Ia and GRBs. In order to access the cosmic history of $\gamma(z)$ our method is applied on two SGL sub-samples defined
 by the velocity dispersions  of lenses\footnote{ {The dynamical mass is related to the velocity dispersion through the relation $M\propto \sigma_{ap}^2$ in the singular isothermal sphere model (Longair 1998).
 Thus, one may consider these sub-samples as being divided by lens masses.}} ($\sigma_{ap}$) and  three SGL sub-samples  defined by  lens redshifts (see section IV for details). Three simple parametrizations
 for $\gamma(z)$ are proposed, namely: (P1) $\gamma(z_l)=\gamma_0 + \gamma_1 z_l$, (P2) $\gamma(z_l)=\gamma_0 + \gamma_1 z_l/(1+z_l)$ and (P3) $\gamma(z)=\gamma_0 + \gamma_1 \ln(1+z_l)$, where $z_l$  corresponds 
to lens redshift. It was obtained a 1$\sigma$ tension on the ($\gamma_0$-$\gamma_1$) plane from the results by
 using the sub-samples with high ($>250$ km/s) and low ($\leq 250$ km/s)  velocity dispersions. On the other hand, the results from the three sub-samples  defined by  lens redshifts are in full agreement
 each other. We also perform analyses with the complete SGL sample. As we shall see, for those accepting the strict validity of the standard CDDR relation, our analyses suggest no significant departure
 from a $\gamma(z_l)$ constant, but a mild evolution is allowed by the data.} 


The paper is organized as follows. In Section 2 we present the methodology, Section 3 contains the data of
strong-lensing used in our analyses, Section 4 presents the analyses and results, and the conclusions are given in Section 5.

\section{Methodology}

In this section we discuss the key aspects of our methodology, such as: the validity of CDDR, SGL observations (Einstein radius, SIS and Power-Law models) and, various parametrizations of $\gamma(z)$.  

\subsection{The cosmic distance duality relation validity}

The main point of our methodology is to consider the validity of the CDDR relation, namely: $D_LD_A^{-1}(1+z)^{-2}=1$. The so-called CDDR is the astronomical version of the reciprocity 
theorem proved long ago by Etherington (1933) and it requires only that source and observer are connected by null geodesics in a Riemannian spacetime and that the number of photons are 
conserved (see also Ellis 1971, 2007).  It plays an essential role in cosmological observations {and has been extensively applied by several authors in different cosmological 
context (Bassett \& Kunz 2004; Cunha, Marassi \& Santos 2007; Zhu et al. 2008; Cao \& Liang 2011; Holanda, Lima \& Ribeiro 2011;
Mantz et al. 2014, Cao et al. 2016b, Rana et al. 2016)}. Recently, several ways to test this relation have been proposed using different astronomical quantities, such as: 
SNe Ia plus $H(z)$ data, gas mass fractions and angular diameter distances of galaxy clusters plus SNe Ia, gamma-ray burst plus $H(z)$, SNe Ia plus barion acoustic oscillations (BAO), 
cosmic microwave background radiation (CMB), gas mass fraction plus $H(z)$ data, SNe Ia plus CMB plus BAO, gravitational lensing plus SNe Ia. An interesting summary with several results
 can be found in Table I of Holanda, Busti \& Alcaniz (2016). As a main conclusion, no significant departure from the validity of the CDDR has been verified.

\subsection{Einstein radius}

An important measurement used in our analyses is the Einstein radius. When the source (s), the observer (o) and the lens (l) in a SGL system are nearly aligned  with each other, then a
 ring like structure is formed called Einstein radius (Schneider, Ehlers \& Falco 1992; Kochanek, Schneider \& Wambsganss 2004). This quantity depends on the evolution of the strong-lensing
 system and  on its mass distribution model. For the simplest one, based on SIS model, the Einstein radius is given by:
\begin{equation}
\theta_E=4\pi {D_{A_{ls}}\over D_{A_{s}}}\frac{\sigma_{SIS}^2}{c^2}\,,\label{thetaE}
\end{equation}
where $\sigma_{SIS}$ is the dispersion velocity due to lens mass distribution, $c$ the speed of light, $D_{A_{ls}}$ and $D_{A_{s}}$ are the angular diameter distances between lens and source, 
and observer and source, respectively. 

As commented early, several studies have shown that slopes of density profiles of individual galaxies exhibit a non-negligible scatter from the SIS model. In this way, the SIS model was
 generalized in order to assume a spherically symmetric power-law mass distribution of type $\rho\sim r^{-\gamma}$ (which becomes a SIS model for $ \gamma= 2$). So, the Einstein radius is written as (Cao et al. 2015a)
\begin{equation}
\theta_E=4\pi {D_{A_{ls}}\over D_{A_{s}}}\frac{\sigma_{ap}^2}{c^2}\bigg({\theta_E\over \theta_{ap}}\bigg)^{2-\gamma}f(\gamma)\,,\label{thetaEgamma}
\end{equation}
where $\sigma_{ap}$ is the stellar velocity dispersion inside an aperture of size $\theta_{ap}$ and
\begin{equation}
f(\gamma)=-\frac{(5-2\gamma)(1-\gamma)}{\sqrt{\pi}(3-\gamma)}\frac{\Gamma(\gamma-1)}{\Gamma(\gamma-3/2)}\bigg[\frac{\Gamma(\gamma/2-1/2)}{\Gamma(\gamma/2)}\bigg]^2\,.
\end{equation}
Therefore\footnote{{A more general expression can be obtained if one relaxes the assumption that the stellar luminosity and total mass distribution follows the same power-law (see Eq. (11) in Cao et al. 2016a).}},
\begin{equation} 
\label{NewObservable}
 D\equiv D_{A_{ls}}/D_{A_{s}} = \frac{c^2 \theta_E }{4 \pi \sigma_{ap}^2} \left( \frac{\theta_{ap}}{\theta_E} \right)^{2-\gamma} f^{-1}(\gamma).
\end{equation}
As we discuss further, such generalization jointly with the CDDR validity allows to study models where the mass profile evolves with redshift, namely $\gamma=\gamma(z)$.
\begin{figure*}
\includegraphics[width=0.3\textwidth]{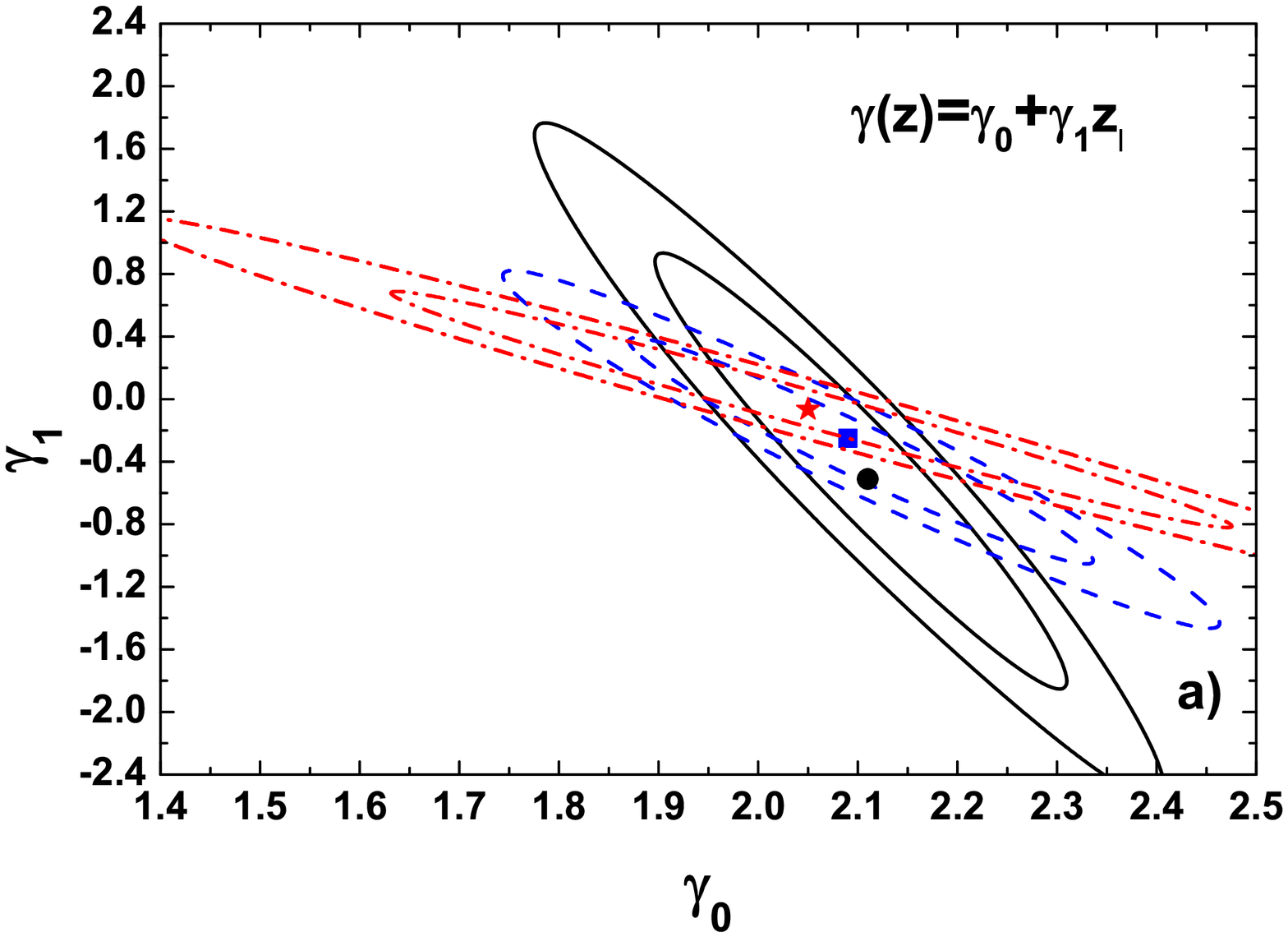}
\includegraphics[width=0.3\textwidth]{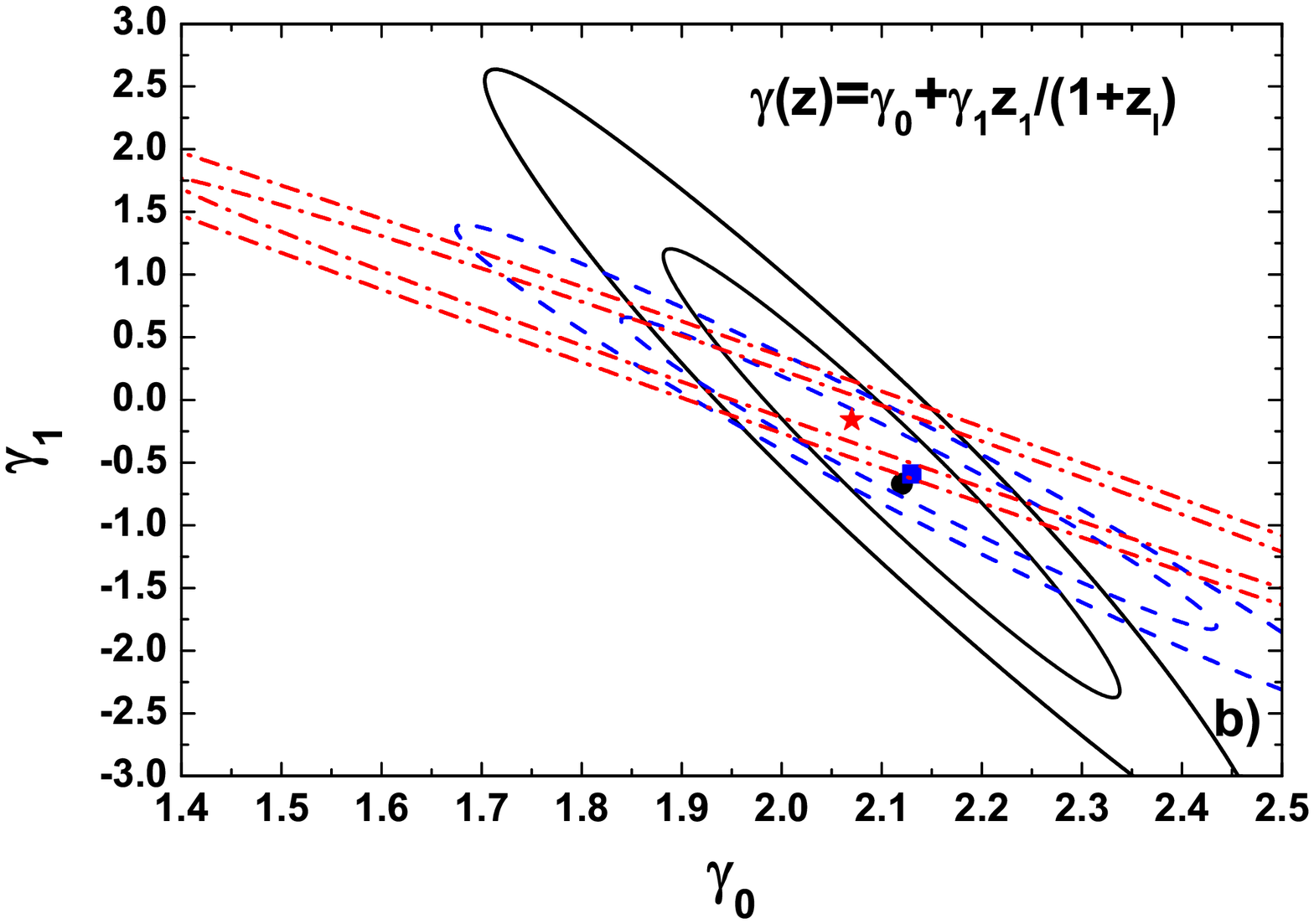}
\includegraphics[width=0.3\textwidth]{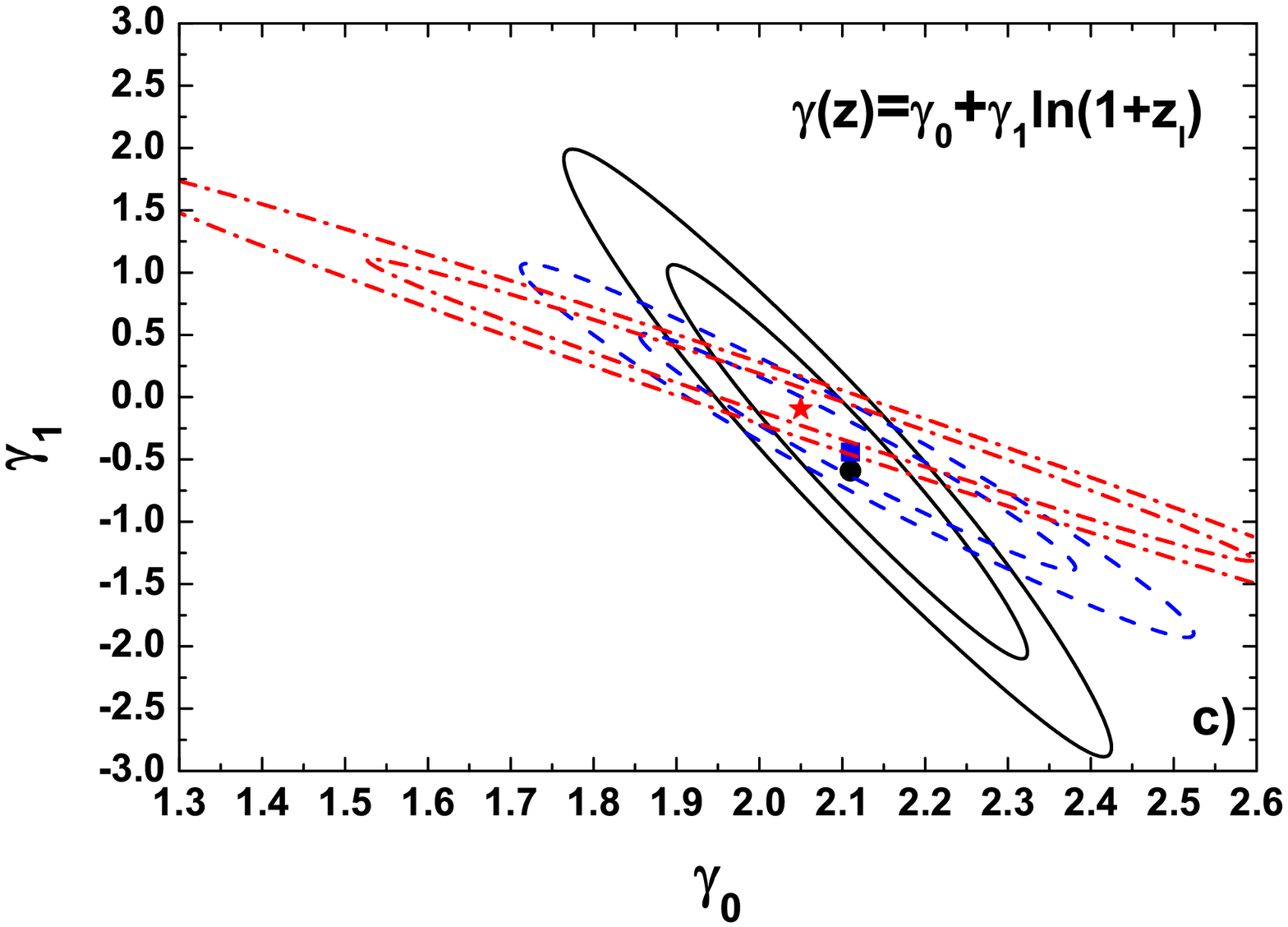}
\caption{Figs. (a), (b)  and (c) show the 1$\sigma$ and 2$\sigma$ {confidence contours in ($\gamma_0$ - $\gamma_1$) plane for all the three parametrizations. Solid black, dashed blue and  dashed-dot red line contours are obtained with the SGL sub-samples having lens redshift: $z_l \leq 0.20$, $0.20 < z_l \leq 0.45$ and $z_l > 0.45$, respectively. The filled red star, blue square and black circle  correspond to best fits for each case.}}
\end{figure*}
\begin{figure*}
\includegraphics[width=0.3\textwidth]{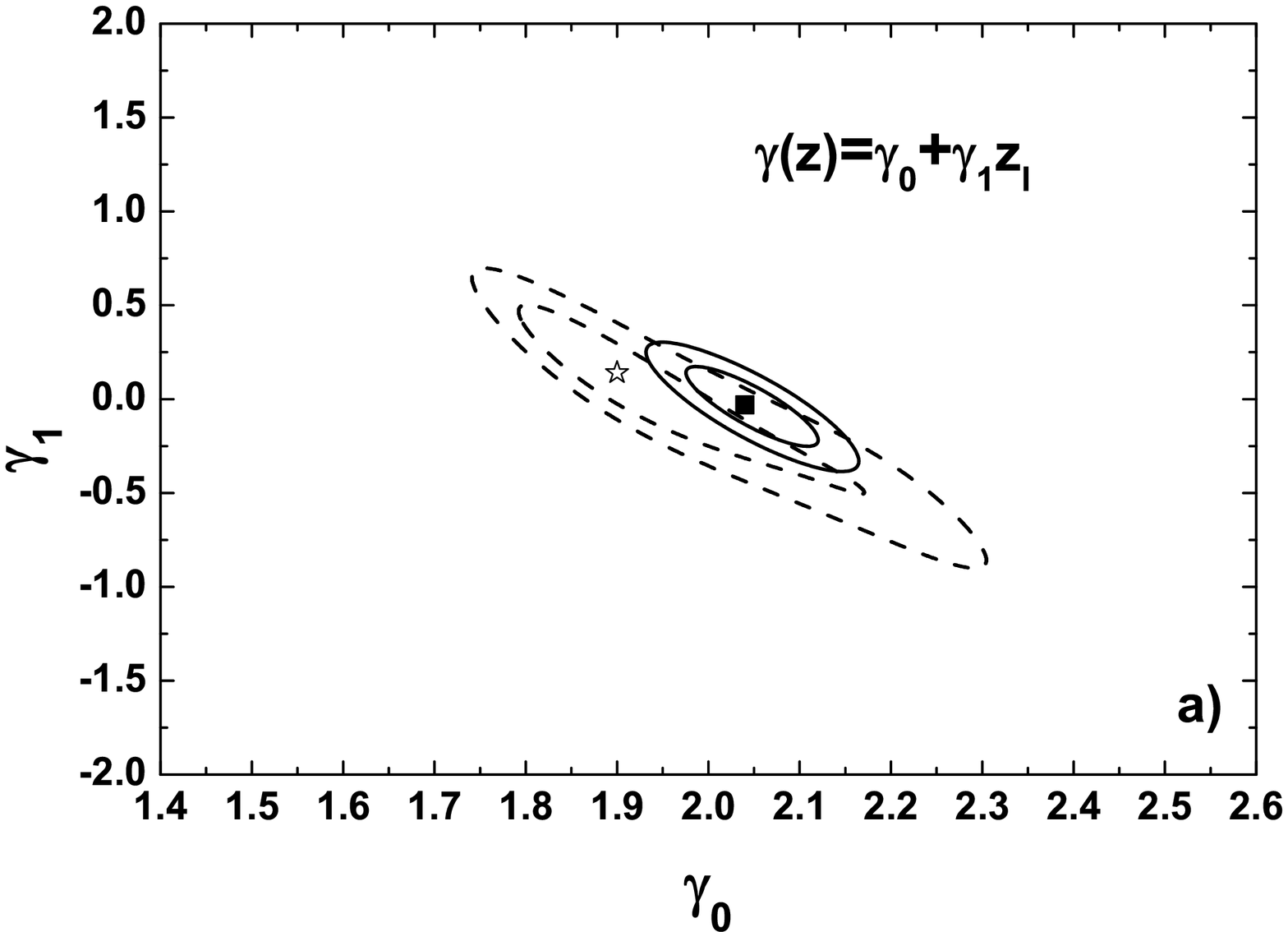}
\includegraphics[width=0.3\textwidth]{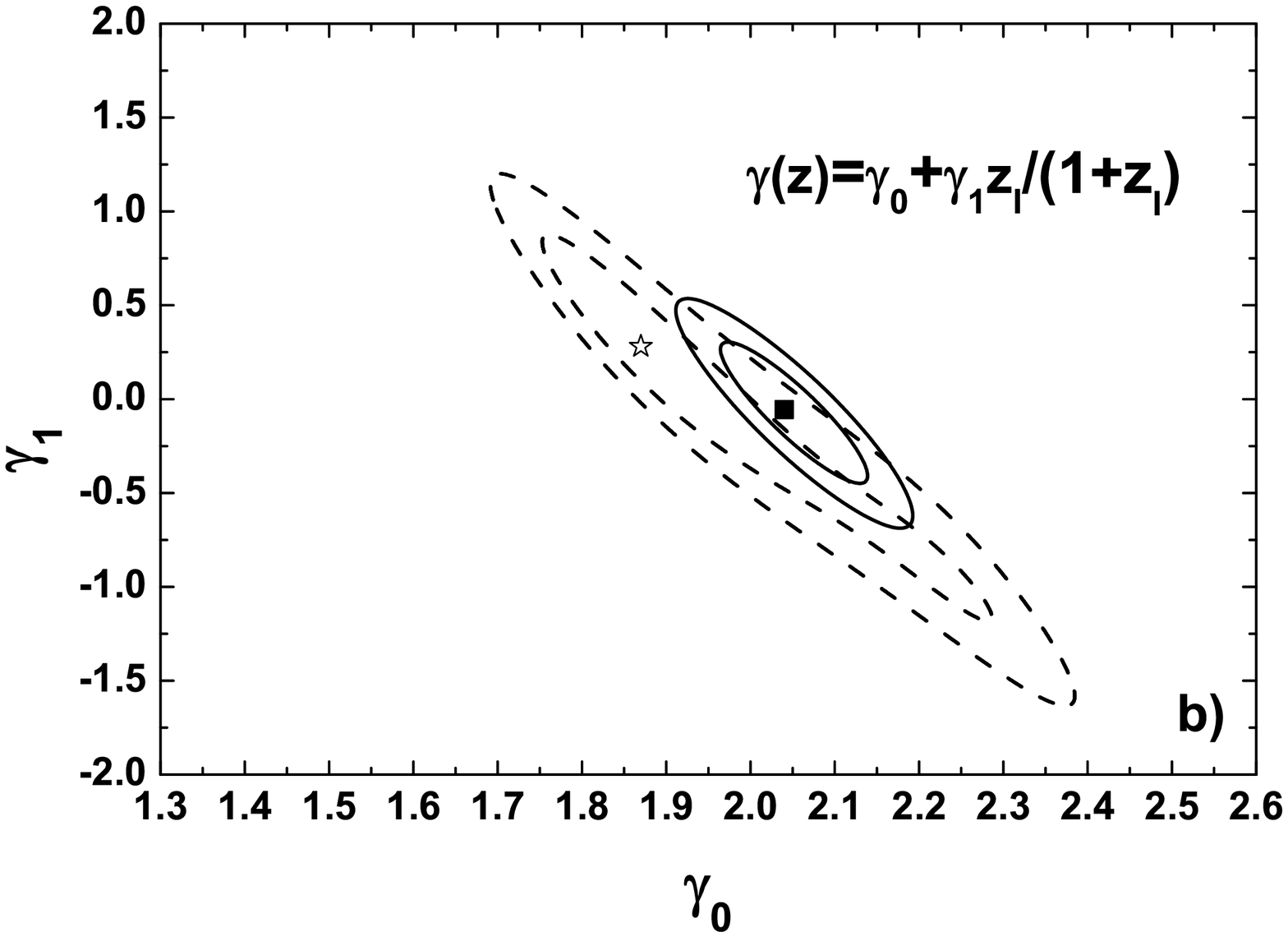}
\includegraphics[width=0.3\textwidth]{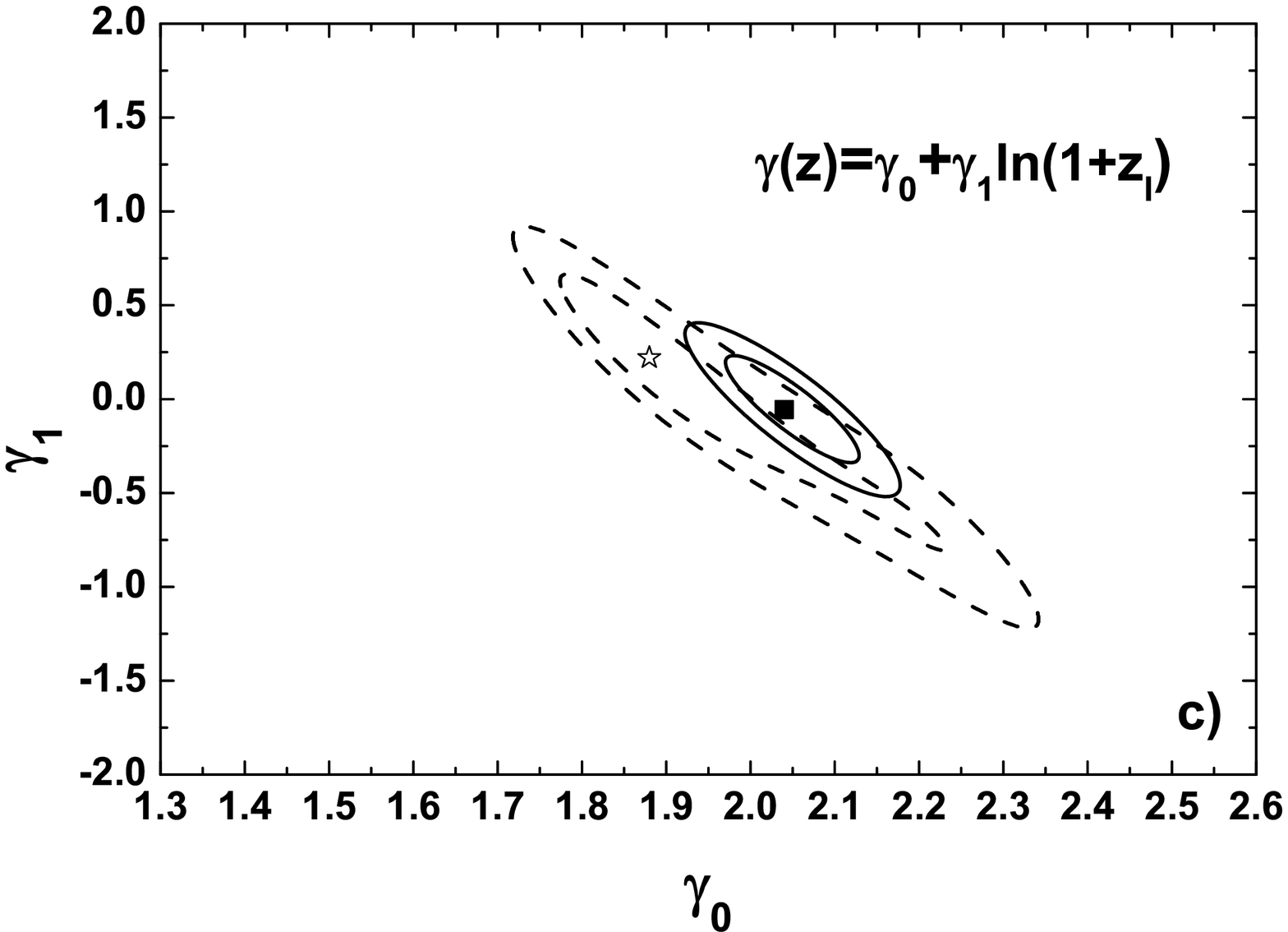}
\caption{Figs. (a), (b)  and (c) show the 1$\sigma$ and 2$\sigma$ {confidence contours in ($\gamma_0$ - $\gamma_1$) plane for all the three parametrizations. Solid black, dashed blue and  dashed-dot red line contours are obtained with the SGL sub-samples with $\sigma_{ap}\leq 250$ km/s  and $ >250$ km/s, respectively.  Open star and filled square  correspond to best fits for each case.} }
\end{figure*}
\begin{figure*}
\includegraphics[width=0.3\textwidth]{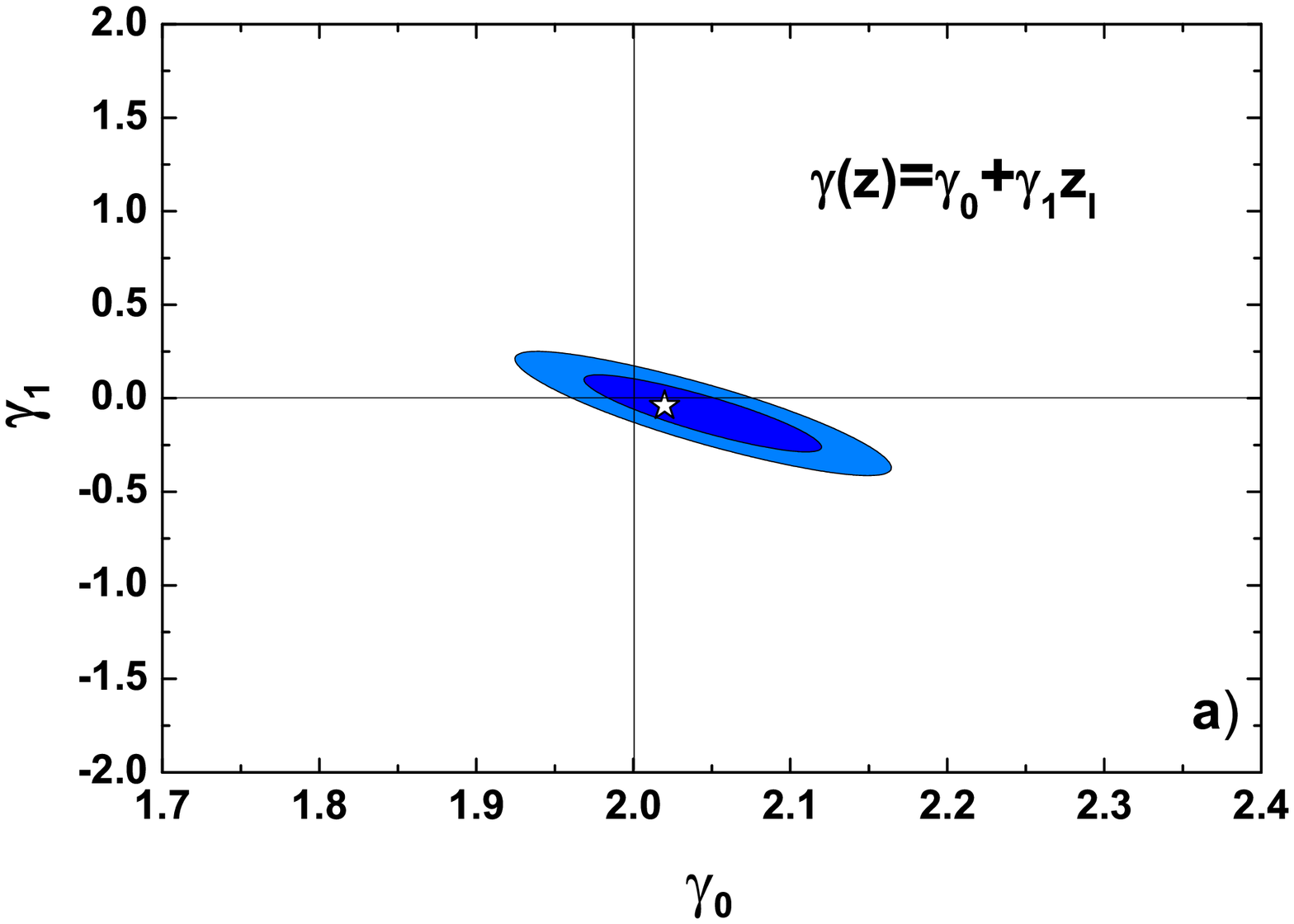}
\includegraphics[width=0.3\textwidth]{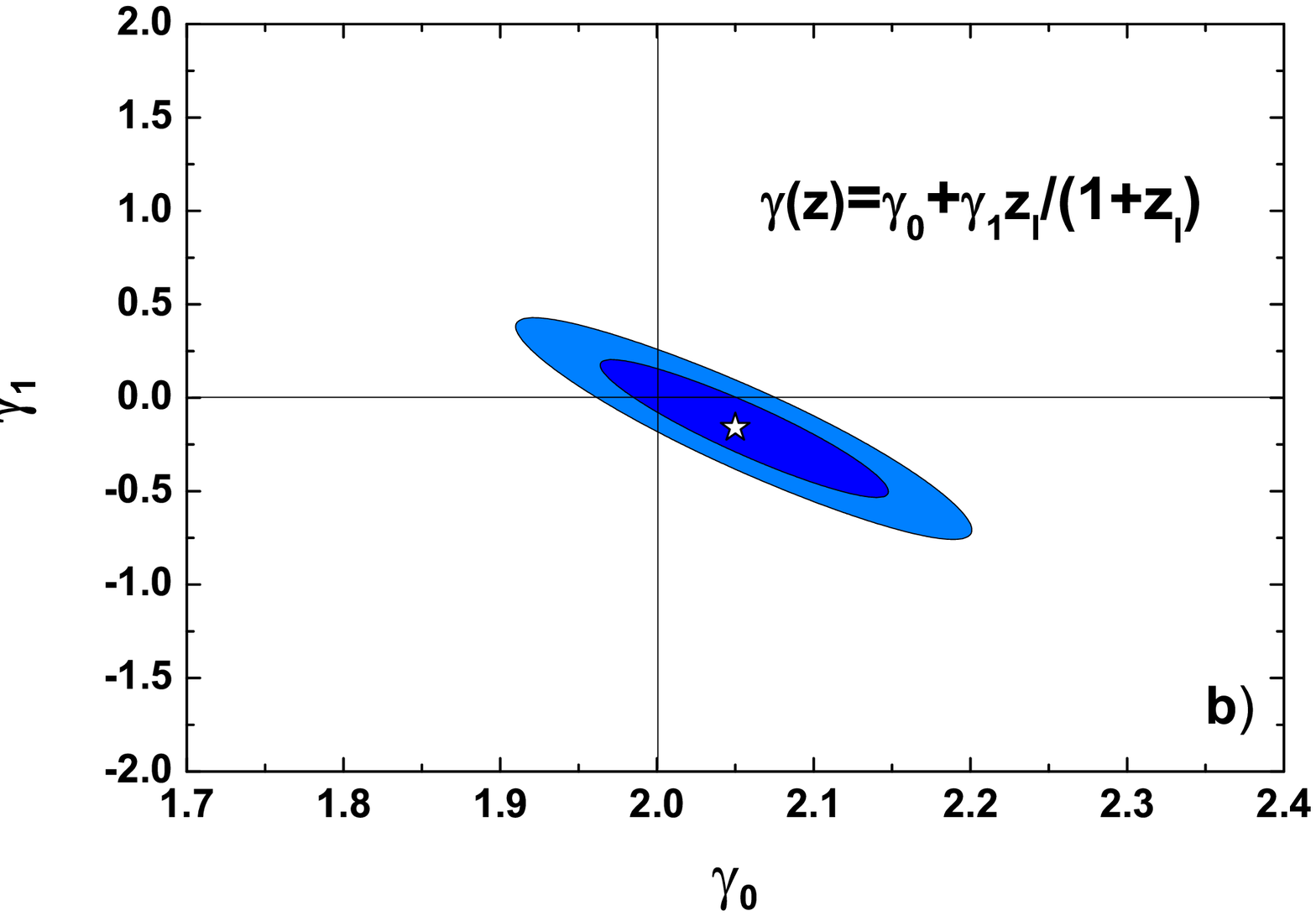}
\includegraphics[width=0.3\textwidth]{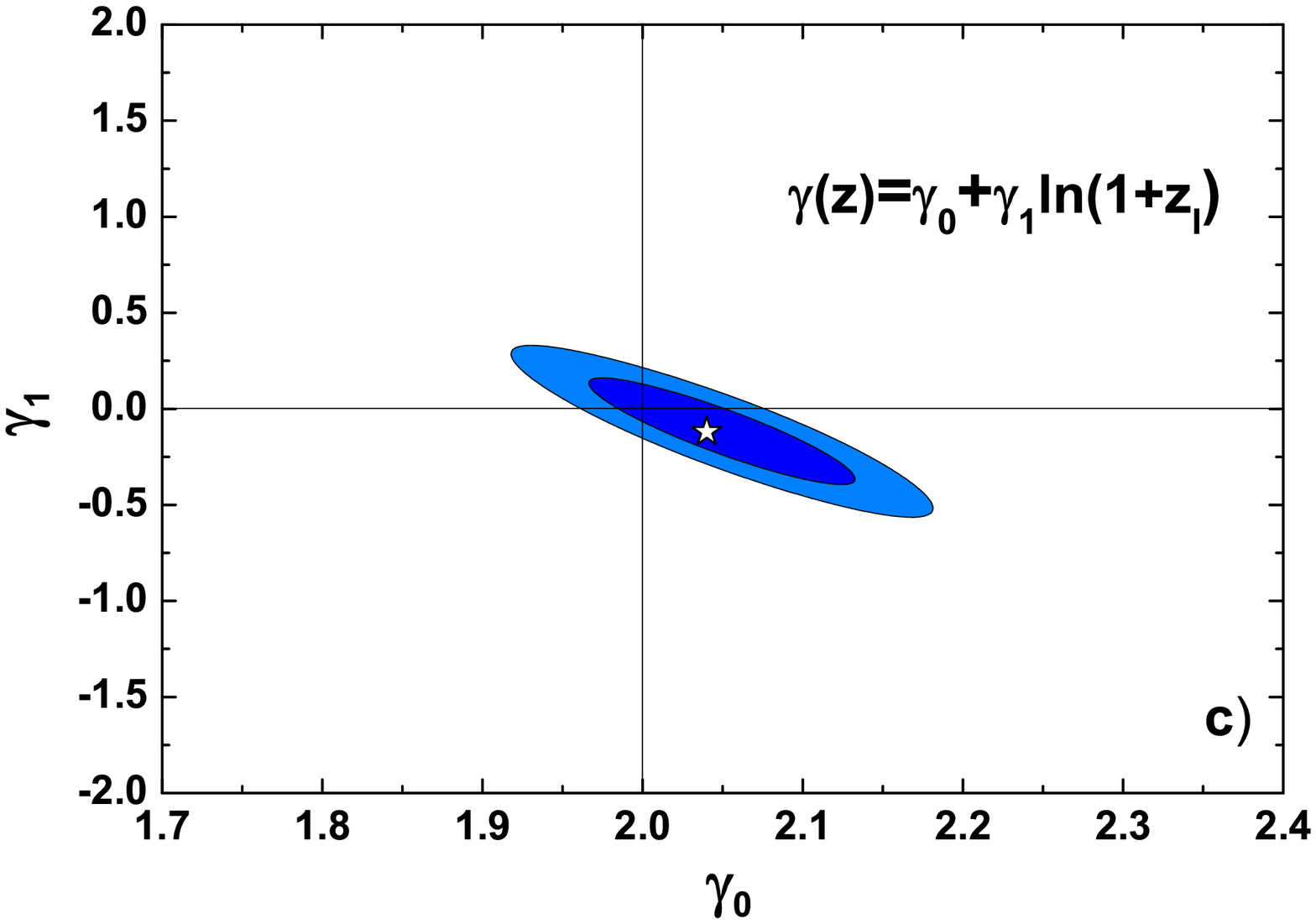}
\caption{ {Figs. (a), (b)  and (c) show the 1$\sigma$ and 2$\sigma$ confidence contours in ($\gamma_0$ - $\gamma_1$) plane for all the three parametrizations with the
complete sample of SGL (92 systems). The white star in each panel corresponds to best fit value.}}
\end{figure*}

\begin{figure*}
\includegraphics[width=0.3\textwidth]{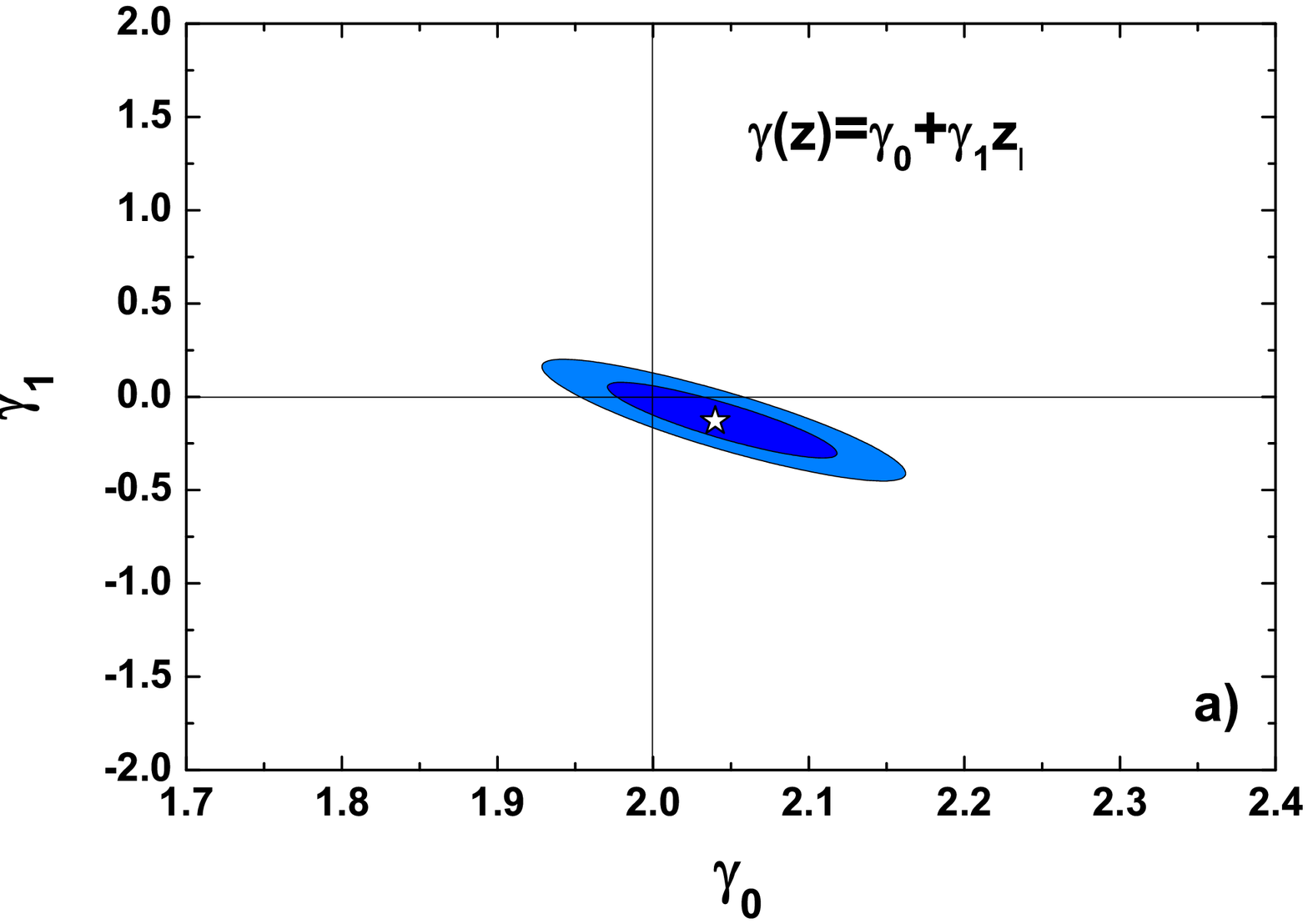}
\includegraphics[width=0.3\textwidth]{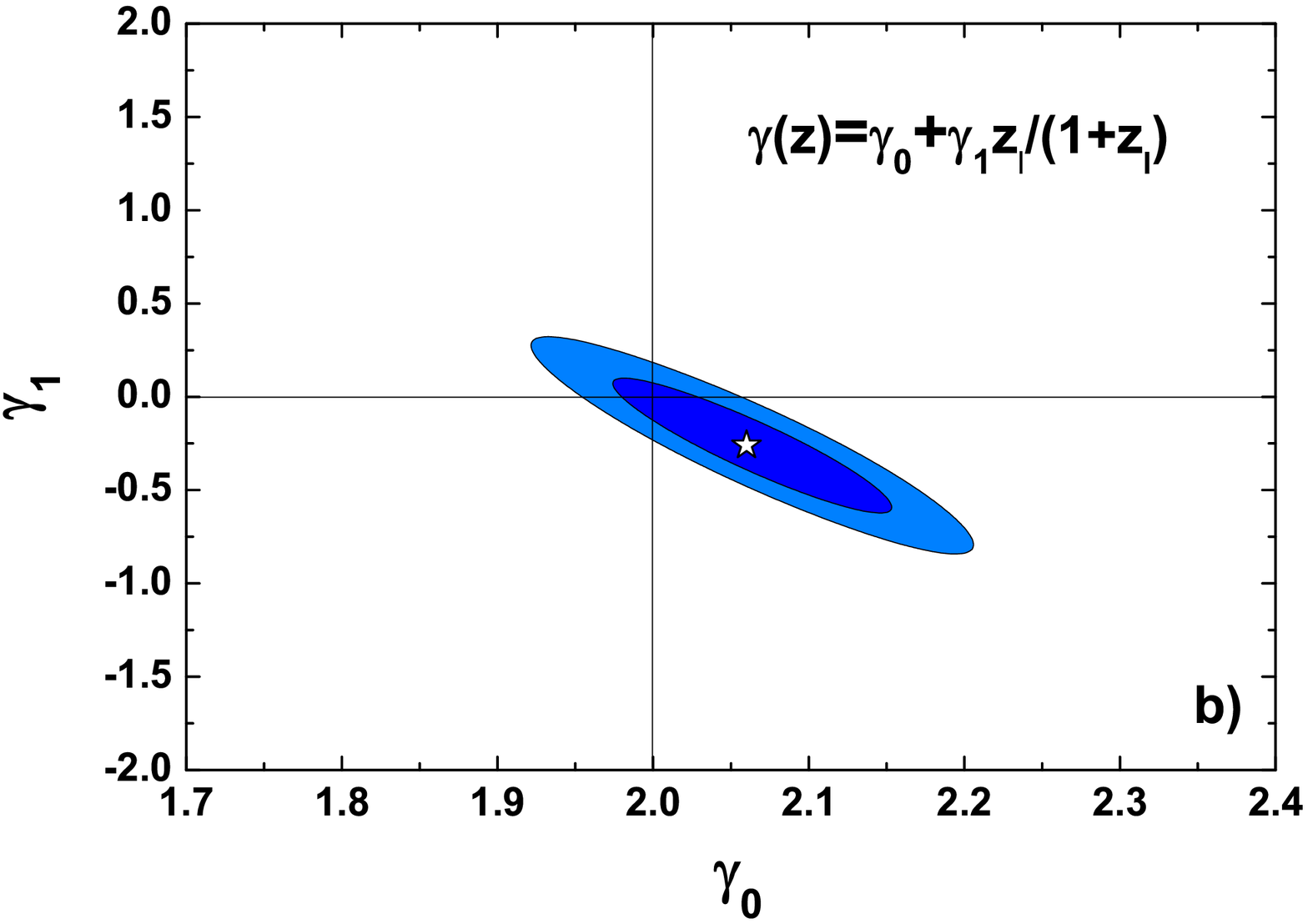}
\includegraphics[width=0.3\textwidth]{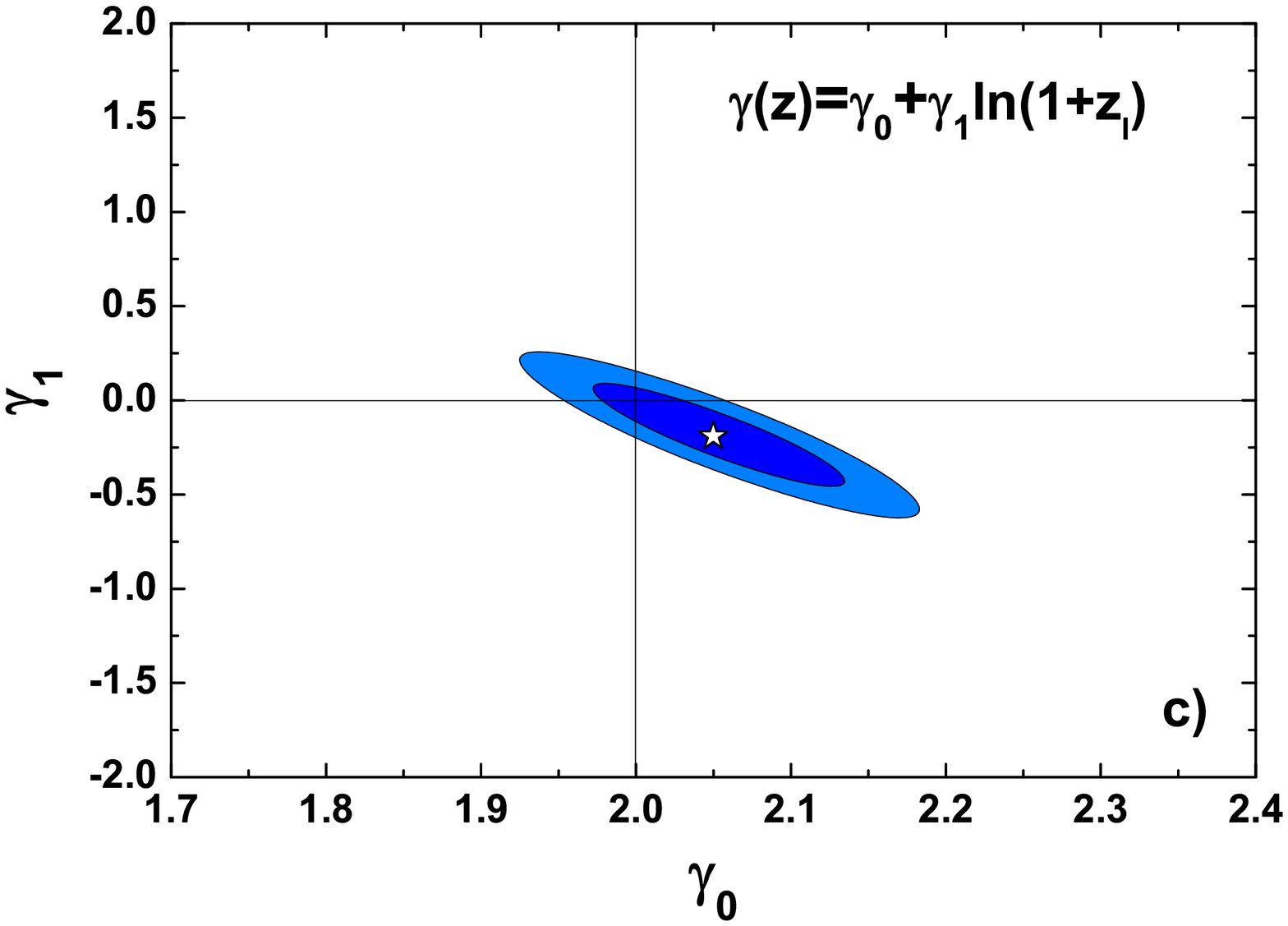}
\caption{ {Figs. (a), (b)  and (c) show the 1$\sigma$ and 2$\sigma$ confidence contours in ($\gamma_0$ - $\gamma_1$) plane for all the three parametrizations by using SNe Ia from JLA compilation (Betoule et al. 2014) plus GRBs and SGL (87 systems). The white star in each panel corresponds to best fit value.}}
\end{figure*}

\subsection{Investigating the $\gamma(z)$ cosmic evolution with the CDDR validity}

Previous papers proposed to test the CDDR validity by using ADD from SGL systems jointly with luminosity distances from SNe Ia data
 (Liao et al. 2016; Holanda, Busti \& Alcaniz 2016; Holanda, Busti, Lima \& Alcaniz 2016). Particularly, in Liao et al. (2016) the method did not depend on  assumptions concerning the
 details of a cosmological model and only a flat universe was assumed. In such flat universe 
the comoving distance between lens and source, $r_{ls}$, is given by (Bartelmann \& Schneider 2001)
\begin{equation}
r_{ls}=r_s-r_l.
\end{equation}
By using the basic definitions $r_s=(1+z_s)D_{A_s}$, $r_l=(1+z_l)D_{A_l}$  and $r_{ls}=(1+z_s)D_{A_{ls}}$, it is possible to find
\begin{equation}
\label{d2}
D= 1 - \frac{(1+z_l)D_{A_{l}}}{(1+z_s)D_{A_{s}}}.
\end{equation}
In our case, we assume the CDDR validity and the above expression can be written as
\begin{equation}
\label{d3}
D=1-\frac{D_{L_l}(1+z_s)}{D_{L_s}(1+z_l)}.
\end{equation}
Thus, $D$ defined as in above equation depends only on luminosity distances, more precisely, on the luminosity distances to lens and sources of SGL systems. In this work,
 these quantities are calculated  by using SNe Ia and GRBs data  (details are given in section 4). On the other hand, the same $D$ can also be calculated by using Eq. (4) from
 the SGL data. In this case, the only unknown factor
is $\gamma(z)$ which can be further parameterized as follows :

\begin{itemize}
\item P1: $\gamma(z_l)=\gamma_0+\gamma_1 z_l$

\item P2: $\gamma(z_l)=\gamma_0+\gamma_1 z_l/(1+z_l)$

\item P3: $\gamma(z_l)=\gamma_0+ \gamma_1 \ln(1+z_l)$.
\end{itemize}
The parametrizations P2 and P3 have not been explored so far in the literature. 

\section{Data}

The following data sets are used in this paper:

\subsection{Angular diameter distances}

\begin{itemize}
\item The original SGL data comprises 118 systems from Cao et al. (2015a) and were observed in the  Sloan Lens ACS survey (SLACS), BOSS Emission-Line Lens Survey (BELLS), Lenses Structure and 
Dynamics Survey (LSD) and Strong Legacy Survey SL2S, with redshift ranges: $0.075 \leq z_l \leq 1.004$ and $0.20 \leq z_s \leq 3.60$.  In Table 1 of Cao et al. (2015a), all relevant information 
necessary to obtain $D$ as defined in Eq. (\ref{NewObservable}) is displayed. 
\end{itemize}

\subsection{Luminosity distances}

\begin{itemize}

\item The  {\bf main} SNe Ia data set used here is taken from Suzuki et al. (2012), which comprises 580 points called Union2.1 compilation.  The SNe Ia redshift 
range is $0.015 \leq z \leq 1.42$. {As is largely known, the distance modulus of Union2.1 compilation was calibrated using the SALT II light curve fitter (Guy et al. 2007). Modern fitters 
as SALT II calibrate cosmological parameters together with light curve parameters. That is, the distance modulus is given by $ \mu = 5 \log (D_L) + 25 = m_B - M + \alpha x - \beta c$, where $M$ is
 the absolute magnitude, $m_B$ is the apparent magnitude, $\alpha$ is the stretch parameter, $\beta$ is the color parameter, and $x$ and $c$ are parameters measured from the light curve. 
No calibration with local objects is performed. The values of distance moduli used in our analyses were calibrated by using an underlying  cosmological model, namely, the flat $\Lambda$CDM. However, as 
the Union2.1 consists of several sub-samples, Suzuki et al. (2012) fit a different $M$ for each sub-sample thereby making the impact of the cosmological model very small (see Section 4.4 of their paper). 
Thus, we believe that the Union2.1 sample is sufficient to turn our analyses  weakly dependent on a specific cosmological model. We also added quadratically a 0.15 magnitude error, which can be associated 
with the intrinsic dispersion of all SNe Ia data.}

\item {Since several sources of SGL systems lie in the interval $1.4 \leq z_s \leq 3.6$, i.e., beyond the redshift range of current SNe Ia compilations ($z\approx 1.50$), we consider also the  
latest GRBs distance modulus data, whose redshift range is $0.033 \leq z \leq 9.3$. The complete sample from Demianski et al. (2017) has 167 GRBs. These authors used a local regression 
technique jointly with SNe Ia luminosity distances (Union2.1) to calibrate  several correlations between spectral and intensity properties, which suggest that GRBs can be used as
 distance indicators. Moreover, no dependence on redshift of the correlations were found.} 
\end{itemize}

\begin{table*}[t]
\caption{Results obtained for the parameters $\gamma_0$ and $\gamma_1$ for each parametrization P1, P2 and P3 in different ranges of redshift $z_l$ and $\sigma_{ap}$ (2 free parameters). All intervals are at 1$\sigma$ c.l..}
\label{tab:example1}
  \begin{tabular}{lcccccc}
    \toprule \hline \\
    \multirow{2}{*}{} &
      \multicolumn{2}{c}{P1} &
      \multicolumn{2}{c}{P2} &
      \multicolumn{2}{c}{P3} \\ 
      & {$\gamma_0$} & {$\gamma_1$} & {$\gamma_0$} & {$\gamma_1$} & {$\gamma_0$} & {$\gamma_1$} \\ \hline \\
      \midrule
    $z_l<0.2$ & $2.11 \pm 0.20$ & $-0.51 \pm 1.32$ & $2.12^{+0.23}_{-0.22}$ & $-0.67^{+1.90}_{-1.83}$ & $2.11^{+0.22}_{-0.24}$ & $-0.59^{+1.97}_{-1.58}$ \\ \\
    $0.2 < z_l < 0.45$ & $2.10 \pm 0.25$ & $-0.33 \pm 0.75$ & $2.13^{+0.33}_{-0.27}$ & $-0.59^{+0.99}_{-1.11}$ & $2.11^{+0.29}_{-0.27}$ & $-0.44^{+0.99}_{-1.05}$ \\ \\
    $z_l > 0.45$ & $2.05 \pm 0.41$ & $-0.065 \pm 0.80$ & $2.07 \pm 0.67$ & $-0.16 \pm 1.95$ & $2.05 \pm 0.54$  & $-0.09^{+1.10}_{-1.03}$ \\ \\
    $\sigma_{ap} < 250$ km/s & $2.04^{+0.08}_{-0.06}$ & $-0.03 \pm 0.23$ & $2.05^{+0.11}_{-0.09}$ & $-0.065 \pm 0.44$ & $2.04^{+0.10}_{-0.09}$ & $-0.05^{+0.31}_{-0.30}$ \\ \\
    $\sigma_{ap} > 250$ km/s & $1.89^{+0.29}_{-0.09}$ & $0.14^{+0.34}_{-0.64}$ & $1.87^{+0.43}_{-0.12}$ & $0.27^{+0.48}_{-1.45}$ & $1.88^{+0.42}_{-0.10}$ & $0.22^{+0.43}_{-0.97}$ \\ \\
    Union2.1+GRB+SGL (92 data points) & $2.04^{+0.08}_{-0.06}$ & $-0.085^{+0.21}_{-0.18}$ & $2.05 \pm 0.10$ & $-0.16^{+0.36}_{-0.34}$ & $2.04 \pm 0.11$ & $-0.12^{+0.29}_{-0.27}$ \\ \\
    JLA+GRB+SGL (87 data points) & $2.04^{+0.08}_{-0.06}$ & $-0.13^{+0.19}_{-0.20}$ & $2.06 \pm 0.10$ & $-0.26^{+0.31}_{-0.35}$ & $2.05 \pm 0.11$ & $-0.19^{+0.26}_{-0.29}$ \\ 
    \bottomrule \hline
  \end{tabular}
\end{table*}

\section{Analyses and Results}

In order to perform the analyses with Eq. (\ref{d3}) we need luminosity distances to the lens and source of each SGL system. These quantities are obtained as follows: {for each one of the 118 SGL systems,
 we carefully select SNe Ia and GRBs with redshifts obeying the criteria\footnote{For 26 SGL systems we do not find SNe Ia (Union2.1) or GRBs obeying the (I) and (II) criteria.} (I) $|z_{l} - z_{SNe/GRB}| \leq 0.006$
 and (II) $|z_{s} - z_{SNe/GRB}| \leq 0.006$. Obviously, the SNe Ia or GRBs obeying (I) and (II) are not the same. Finally, we calculate the following weighted average for the distance modulus selected in each case}
\begin{equation}
\begin{array}{l}
\bar{\mu}=\frac{\sum\left(\mu_{i}/\sigma^2_{\mu_{i}}\right)}{\sum1/\sigma^2_{\mu_{i}}} ,\hspace{0.5cm}
\sigma^2_{\bar{\mu}}=\frac{1}{\sum1/\sigma^2_{\mu_{i}}}.
\end{array}\label{eq:dlsigdl}
\end{equation}
After all, we end with a sample containing 92 SGL systems (see Fig. 1a) and 184 $\bar{\mu}$ from SNe Ia (Union2.1) and GRBs data (two $\bar{\mu}$  for each SGL system are necessary).
 Naturally, $\bar{D}_L=10^{(\bar{\mu}-25)/5}$ and $\sigma^{2}_{\bar{D}_L}=\left(\frac{\partial \bar{D}_L}{\partial \bar{\mu}}\right)^2\sigma^{2}_{\bar{\mu}}$ (see Figs. 1b and 1c).
 { Considering only SNe Ia data (Union2.1), the final sample would have only 65 SGL systems.}

The constraints on the $\gamma_0$ and $\gamma_1$ parameters are obtained by evaluating the likelihood distribution 
function, ${\cal{L}} \propto e^{-\chi^{2}/2}$, with

\begin{eqnarray}
\chi^2 &=& \sum_{i=1}^{92}\frac{\left[D_i(\gamma_0,\gamma_1)-1+\frac{\bar{D}_{iL_l}(1+z_s)}{\bar{D}_{iL_s}(1+z_l)})\right]^2}{\sigma_{iobs}^2}\nonumber,
\label{chi}
\end{eqnarray}
where $D$ is given by Eq.(4), which depends on $\gamma_0$ and $\gamma_1$, and $\sigma_{iobs}^2$ stands for the statistical errors associated to the $D_L(z)$ from SNe Ia 
and GRBs data and to gravitational lensing  observations. The $\sigma_D$ error is given by
\begin{equation} \label{uncertainty}
\sigma_D = D \sqrt{4 (\delta \sigma_{0})^2 + (1-\gamma)^2 (\delta \theta_E)^2}\;.
\end{equation}

As discussed earlier, the statistical analyses are performed considering six SGL sub-samples, namely,

\begin{itemize}

\item 53 SGL systems with $\sigma_{ap} \leq 250$ km/s (low-intermediate lens masses)
\item 39 SGL systems with $\sigma_{ap} > 250$ km/s (intermediate-high lens masses)
\item 25 SGL systems with $z_{l} \leq 0.20$ (low redshifts)
\item 44 SGL systems with $0.20 < z_{l}\leq 0.45$  (intermediate redshifts)
\item 23 SGL systems with $ z_{l} > 0.45$  (high redshifts)
\item  Complete sample (92 SGL systems) obtained by using Union2.1 SNe Ia + GRBs

\end{itemize}
{As commented by Cao et al. (2016a), elliptical galaxies with velocity dispersion smaller than 200 km/s may be classified roughly as relatively low-mass galaxies,  
while those with velocity dispersion larger than 300 km/s may be treated as relatively high-mass galaxies. Naturally, elliptical galaxies with velocity dispersion 
between 200-300 km/s may be classified as intermediate-mass galaxies. In order to guarantee that there is enough data in each sub-sample, we consider only two sub-samples when 
the velocity dispersion is used as criterion.}

{Our results are plotted in Figures (2), (3), (4) and best fit values are mentioned in Table 1. Figures (2a), (2b) and (2c) show the 1$\sigma$ and 2$\sigma$ confidence 
regions in the ($\gamma_0$-$\gamma_1$) plane considering the three $\gamma(z_l)$ parametrizations and SGL sub-samples defined by different lens redshifts. In each panel, 
results obtained with the SGL sub-samples with $z_l \leq 0.20$, $0.20 < z_l \leq 0.45$ and $z_l > 0.45$ are shown with solid black, dashed blue and  dashed-dot red lines, respectively. 
The filled red star, blue square and black circle  correspond to the best fits for each case, respectively. From Table 1, one may see that the values are in full agreement each other 
and the P1 parametrization gives the more restrictive intervals. In all cases, the SGL sub-samples in low and intermediate redshifts provide tighter  regions in parameter space. 
The best fits of the $\gamma_1$ in parametrizations by using the SGL sub-sample with $z_l > 0.45$ are closer to zero than the other SGL sub-samples, but in all cases the central 
value is negative, suggesting a slight evolution to $\gamma(z_l)$.}

{Figures (3a), (3b) and (3c) show the 1$\sigma$ and 2$\sigma$ confidence regions for the ($\gamma_0$-$\gamma_1$) plane considering the three $\gamma(z_l)$ parametrizations and SGL 
sub-samples defined by different  velocity dispersions of lenses.  The SGL sub-samples with $\sigma_{ap} \leq 250$ km/s, $\sigma_{ap} > 250$ km/s are represented by the solid and 
dashed black lines, respectively. The filled black square and the open star  correspond to the best fits for each case. Again, for each sub-sample, the regions in parameter space 
depend weakly  on the $\gamma(z_l)$ parametrization. However, by comparing the contours obtained with the SGL sub-samples in each panel, one may see that the  1$\sigma$ regions for ($\gamma_0$-$\gamma_1$) 
are incompatible with each 
other. Moreover, in all cases, the best fits of the $\gamma(z)$ parametrizations by using the SGL sub-sample with $\sigma_{ap} > 250$ km/s are rule out in 2$\sigma$ c.l. by 
the confidence regions of the SGL sub-sample with $\sigma_{ap} \leq 250$ km/s. Finally, the best fits of $\gamma_1$ are always positive when $\sigma_{ap} > 250$ km/s, while for 
the other sub-sample are negative. These results show a interesting dependence of the $\gamma$ parameter on the mass lens. } 

 {Figures (4a), (4b) and (4c) show the 1$\sigma$ and 2$\sigma$ confidence regions for the ($\gamma_0$-$\gamma_1$) plane considering the 92 SGL systems, the respective
 luminosity distances and the three $\gamma(z)$ parametrizations. The open star corresponds to the best fits. As one may see in Table 1, we
 obtain $\gamma_0 \approx 2$ and $ \gamma_1 \approx 0$ within 1$\sigma$ c.l. for all $\gamma(z_l)$ parametrizations. However, the best fit values of $\gamma_1$  are slightly
 negatives: $-0.085$, $-0.16$ and $-0.12$ for P1, P2 and P3 parametrizations, respectively, suggesting a mild evolution for $\gamma(z)$.  }

  { We also perform a analysis by using the SNe Ia from JLA compilation (Betoule et al. 2014) plus GRBs. For this case,  we obtain a sub-sample with 87 
SGL and the respective luminosity distances from JLA and GRBs. The JLA compilation contain 740 spectroscopically confirmed SNe Ia in the redshift range of $ 0.01< z < 1.3$. The distance modulus  of each SNe Ia  further depend  upon on the $ \alpha$,$ \beta$ and $M$ as mentioned in section 3.2 for Union2.1 compilation. In  recent works (Nielsen, Guffanti \&  Sarkar 2016; Evslin 2016; Nunes, Pan \& Saridakis 2016; Li et al. 2016b), it has been observed that $\alpha$ and $\beta$ 
 act like a global parameters,  whatever prior cosmological model one choose to find the distance modulus of each SNe. So we fix the values of $\alpha$ and $\beta$ as given by Betoule at al. (2014).
 The results obtained with this dataset ( 87 data points) are plotted in 
 Figures (5a), (5b) and (5c). They show the  1$\sigma$ and 2$\sigma$ confidence 
regions in  the ($\gamma_0$-$\gamma_1$) plane for all the three $\gamma(z)$ parametrizations. The open star corresponds to the best fits. As one may see in Table 1 (last line), we obtain 
again $\gamma_0 \approx 2$ and $ \gamma_1 \approx 0$ within 1$\sigma$ c.l. for all $\gamma(z_l)$ parametrizations. The 1$\sigma$ and 2$\sigma$ c.l. regions are in full agreement 
with those from Union2.1 plus GRBs (fig.4).  However, the best fit values of $\gamma_1$  are more negatives, we obtain: $-0.13$, $-0.26$ and $-0.19$ for P1, P2 and P3 parametrizations, respectively, 
reinforcing a mild evolution for $\gamma(z)$. }

\subsection{Comparing results}

 {It is interesting to compare our results by using the 92 and 87 SGL systems (last two lines in Table 1) with previous ones where $\gamma_0$ and $\gamma_1$ were 
constrained  by adopting the P1 parametrization and different cosmological model in analyses.} For instance:

\begin{itemize}
\item Cao et al. (2015a) used exclusively 118 SGL systems and they found $\gamma_0 = 2.13^{+0.07}_{-0.12} $ and $\gamma_1=-0.09 \pm 0.17$ in a $\omega$CDM model, and $\gamma_0 = 2.14^{+0.07}_{-0.10}$ 
and $\gamma_1=-0.10 \pm 0.18$ in a $\omega(z)$CDM (CPL model) framework. In both cases the matter density parameter was fixed ($\Omega_M=0.315$) based on the Planck results (Ade et al. 2014). 
\item Cui, Li \& Zhang (2017), more recently, by using the SGL observations in combination with other cosmological observations (BAO, CMB and $H(z)$ data),  considered some 
simple dark energy models, such as $\omega$CDM, the holographic dark energy model (Li 2004) and the Ricci dark energy  model (Gao et al. 2009). Briefly, these 
authors derived $\gamma_0 \approx 2.10$ (with the uncertainty around $0.04$-$0.05$) and $\gamma_1 \approx −0.06$ (with the uncertainty around 0.1). 
\item Li et al. (2016a) considered SGL observations plus BAO measurements and found the following values: $\gamma_0 = 2.094^{+0.053}_{-0.056} $ and $\gamma_1=-0.053^{+0.103}_{-0.102}$ in a $\Lambda$CDM 
model, $\gamma_0 = 2.088^{+0.055}_{-0.056} $ and $\gamma_1=-0.054^{+0.104}_{-0.02}$ in a $\omega$CDM model, $\gamma_0 = 2.087^{+0.055}_{-0.056}$ and $\gamma_1=-0.055 \pm 0.105$ in a $\omega(z)$CDM 
(CPL model), $\gamma_0 = 2.087^{+0.052}_{-0.054} $ and $\gamma_1=-0.052^{+0.104}_{-0.102}$ in 
a  Ricci dark energy  model and, finally, $\gamma_0 = 2.074^{+0.050}_{-0.051} $ and $\gamma_1=-0.047^{+0.101}_{-0.102}$ in a Dvali-Gabadadze-Porrati  brane world model.
\end{itemize}
{As one may see, these previous results for $\gamma_0$ are in agreement with the present work within 1$\sigma$ c.l., although they show a  departure from $\gamma_0=2$ at least at 1$\sigma$ c.l.. On the 
other hand, in all cases $\gamma_1 \approx 0$ is allowed  within 1$\sigma$ c.l., indicating that a significant time-varying $\gamma$ is not supported by the current observations.}

{ Finally, we also compare our results with those from Cao et al. (2016a). These authors used the complete SGL sample (118 points)  from Cao et al. 2015a, the flat $\Lambda$CDM 
model (Ade et al. 2014), the P1 parametrization and 6 sub-samples similar to those considered in the present work.  The sub-samples consists of: 25, 80 and 13 SGL 
systems with $\sigma_{ap} \leq 200$ km/s, $200< \sigma_{ap} \leq 300$ km/s and $\sigma_{ap} > 300$ km/s, respectively, and, 25, 65 and 80 SGL systems with $z_{l} \leq 0.20$, $0.20 < z_{l} \leq 0.50$ 
and $ z_{l} > 0.50$, respectively. The main points are :
\begin{itemize}
\item Results obtained by using the 118 SGL systems ($\gamma_0=2.132 \pm 0.055$, $\gamma_1=-0.067 \pm 0.119$) are in full agreement with the present work when the 
92 SGL systems are used, but their result for $\gamma_0$ value is incompatible with 2.0 at least for 1$\sigma$ c.l.. The$\gamma_1$ value is in full agreement with ours (within 1$\sigma$ c.l.).
\item Cao et al. (2016a) results from the sub-samples defined by the different lens redshifts are in full agreement with ours (within 1$\sigma$ c.l.).
\item By using the lens velocity dispersions as criterion, similar behavior for $\gamma(z)$ is found if one compares our SGL sub-sample of low velocity dispersions 
with their of low and intermediate velocity dispersions. However, we obtain for the $\gamma_1$ parameter  a best fit value more positive (Cao et al. 2016a found for this case $\gamma_1=-0.047$).
 The source of this difference may lie in the samples used in the analyses or in the cosmological model considered. 
\end{itemize}
Naturally, our error bars are larger since we have performed analyzes without using a specific cosmological model.}

\section{Conclusion}

Knowing the exact profile of mass distribution for strong gravitational lensing systems is very important in order to use this phenomenon as a precise cosmological tool. In analyses with 
time-delay distance, for instance, different assumptions lead to  different $H_0$ estimates. The simplest model used frequently in strong gravitational lensing observations is the singular
 isothermal sphere (the SIS model). However, it has been changed by a power-law mass distribution ($\rho \propto r^{-\gamma}$) since recent studies in elliptical galaxies have shown  a
 non-negligible scatter from the SIS model. A crucial point in the power-law mass distribution is to know if the $\gamma$ parameter varies with redshift, since this fact is linked to massive galaxies growth process.

 In this paper we propose  a new method  to access a possible $\gamma$  variation. Our theoretical framework was based on two assumptions: a flat universe and the validity of cosmic distance duality
 relation. No specific cosmological model was used. We also considered three $\gamma(z)$ parametrizations, namely: (P1) $\gamma(z_l)=\gamma_0+\gamma_1 z_l$, (P2) $\gamma(z_l)=\gamma_0+\gamma_1 z_l/(1+z_l)$ 
and (P3) $\gamma(z_l)=\gamma_0+\gamma_1 \ln(1+z_l)$. By using 92 strong gravitational lensing observations plus SNe Ia (Union2.1) and GRBs we find no significant $\gamma(z)$ evolution. However, in all cases the best fit 
values for the $\gamma_1$ parameter were found to be negative (except in the sub-sample $\sigma_{ap}>250$ km/s), indicating a mild evolution for $\gamma(z_l)$. Although less restrictive, our results 
are also in full agreement with recent results
 from other cosmological model dependent methods (see Section 4). The lenses and sources of the SGL systems  lie in the redshift range $ 0.073 \leq z_l \leq 0.783 $ and $ 0.0196 \leq z_s \leq 3.59 $. 
 {The mild evolution was reinforced when we considered  a sub-sample with 87 SGL systems and the respective luminosity distances obtained from the JLA SNe Ia compilation and GRBs.}

 {We also considered the  analyses by using sub-samples of the SGL systems defined by different lens redshifts and velocity dispersions plus SNe Ia (Union2.1) and GRBs. }The
 results obtained from sub-samples with $z_l \leq 0.20$, $0.20 < z_l \leq 0.45$ and $z_l > 0.45$ (where $z_l$ is the lens redshift)  are  in full agreement each other. On the other hand,  we found 
that the best fits for the SGL sub-sample with $\sigma_{ap} > 250$ km/s are ruled out in 2$\sigma$ c.l. by the confidence regions of the SGL sub-sample with $\sigma_{ap} \leq 250$. Moreover, the best
 fits of the SGL sub-sample with $\sigma_{ap} < 250$ km/s are negative, while for the other SGL sub-sample are positive. Our results reinforce the need of treating galaxies with low and high velocity dispersions separately.

In the near future, it is expected that several surveys (EUCLID mission, Pan-STARRS, LSST, JDEM) discover thousands of strong lensing systems. Then by applying this method along with bigger sample,
 more stringent limits on the parameters $\gamma_0$ and $\gamma_1$ can be obtained. Besides, as an  interesting extension of the present paper, one may check  the consequences of relaxing the rigid assumption
 that the stellar luminosity and total mass distributions follow the same power law. Also, it would
 be interesting in the future to apply this method with the inclusion of other sources at  cosmological distances, such as powerful radio sources  {(Gurvitz 1994, Gurvitz, Kellermann \& Frey 1999, 
Jackson 2004 Jackson \& Jannetta 2006, Cao et al. 2015b, 2017b)}.
The inclusion of these high redshift sources further  would corroborate or even contradict our present conclusions.

\section*{Acknowledgments}
RFLH acknowledges financial support from  {Conselho Nacional de Desenvolvimento Cient\'ifico e Tecnol\'ogico} (CNPq) and UEPB (No. 478524/2013-7, 303734/2014-0). SHP is 
grateful to CNPq, for financial support (No. 304297/2015-1 and 400924/2016-1). The authors are grateful to referee for very constructive comments.


\end{document}